\documentclass[useAMS,usenatbib]{mn2e}
\usepackage{graphicx}

\usepackage{subfigure}
\usepackage{amsmath}
\usepackage{mathabx}

\usepackage{subfigure}
\usepackage{amsmath}

\def\spose#1{\hbox to 0pt{#1\hss}}
\def\lta{\mathrel{\spose{\lower 3pt\hbox{$\mathchar"218$}}
     \raise 2.0pt\hbox{$\mathchar"13C$}}}
\def\gta{\mathrel{\spose{\lower 3pt\hbox{$\mathchar"218$}}
     \raise 2.0pt\hbox{$\mathchar"13E$}}}

\title[Simulating Disk Galaxies]{MaGICC Disks:  Matching Observed Galaxy Relationships Over a Wide Stellar Mass Range}
\author[C. B. Brook et al.]{C. B. Brook$^{1,2}$, G. Stinson$^{3}$, B. K. Gibson$^{2}$, J. Wadsley$^{4}$, T. Quinn$^{5}$\\
        $^1$Departamento de F\'{i}sica Te\'{o}rica, Universidad Aut\'{o}noma de Madrid, E-28049 Cantoblanco, Madrid, Spain\\ 
   $^2$Jeremiah Horrocks    Institute, 
            University of Central Lancashire, 
            Preston, PR1~2HE, UK \\
$^3$Max-Planck-Institut f\"ur Astronomie, K\"onigstuhl 17, 69117
  Heidelberg, Germany   \\     
            $^4$Department of Physics and Astronomy, McMaster University, Hamilton, ON L8S 4M1, Canada\\
        $^5$Astronomy Department, University of Washington, 
            Box 351580, Seattle, WA 98195-1580, USA \\                  
                }

\begin{document}

\date{}

\pagerange{\pageref{firstpage}--\pageref{lastpage}} \pubyear{2011}

\maketitle

\label{firstpage}

\begin{abstract}
We use the same physical model to simulate four galaxies that match the 
relation between stellar and total mass, over a mass range that includes 
the vast majority of disc galaxies. The resultant galaxies, part of the 
Making Galaxies in a Cosmological Context (MaGICC) program, also match 
observed relations between luminosity, rotation velocity, size, colour, 
star formation rate, HI mass, baryonic mass, and metallicity. Radiation 
energy feedback from massive stars and supernova energy  balance the complex interplay between 
cooling gas, regulated star formation, large scale outflows, and recycling of gas 
in a manner which correctly scales with the mass of the galaxy. Outflows,  driven by the expansion of shells and 
superbubbles of overlapping supernova explosions, 
also play a key role in simulating galaxies with exponential surface 
brightness profiles, flat rotation curves and dark matter cores.   Our study implies that large-scale outflows are the 
primary driver of the dependence of disc galaxy properties on mass. We 
show that the degree of outflows invoked in our model is required to 
meet the constraints provided by observations of OVI absorption lines in 
the circum-galactic media of nearby galaxies.
  \end{abstract}


\section{Introduction}
The 
rotation velocity, size, luminosity \cite[e.g.][]{courteau07}, star 
formation rate \citep[e.g.][]{salim07}, stellar mass \citep[e.g.][]{bell03}, total 
virial mass (including dark matter, \citealt[e.g.][]{mandelbaum06,more11}), mass 
of neutral hydrogen gas \citep[e.g.][]{verheijen01}, total baryon mass 
\citep[e.g.][]{mcgaugh05} and the abundance of oxygen relative to hydrogen 
\citep[e.g.][]{tremonti04} are observationally well-determined. Together, the 
relation between these properties provide stringent constraints which 
simulations of galaxy formation must satisfy.

A large number of such  disc galaxy properties vary with mass.  Growing evidence suggests that the ratio of 
stellar mass to total mass of galaxies is low compared to the ratio of 
baryons to dark matter in the Universe. The relation is steep in the 
mass range where most disc galaxies exist, below a stellar mass $\sim$2$\times10^{10}$M$_\odot$, where a 
doubling of halo mass results in a factor of eight more stellar mass.
The relation has been 
determined by rank-ordering the masses of dark matter halos formed in 
simulations within the cold dark matter cosmological framework, and then 
rank-ordering galaxies based on their stellar mass as determined in 
large scale galaxy surveys, and directly matching stellar and halo 
masses, assuming a monotonic relation between the two 
\citep{moster10,guo10}. The results agree remarkably well with direct 
derivations of the relation, where dark matter halo masses are 
determined by gravitational lensing \citep{mandelbaum06} and the 
dynamics of satellite galaxies \citep{more11}.

Recent progress in simulating   galaxies in a cosmological cold dark matter framework has resulted in the ability to match  an increasing number of properties of observed disk galaxies \citep{governato10,piontek11,agertz11,guedes11} but have not been able to model the dependence of galaxy properties on mass.  

Two approaches have emerged to regulate star formation in simulations.  One method is to lower star formation efficiency either directly \citep{agertz11} or using small scale star formation physics based on the availability of molecular hydrogen (H$_2$) from which stars form \citep{kuhlen11}.  \citet{agertz11}  produced a simulated galaxy which shares many features with the Milky Way, while \citep{kuhlen11} produced low mass galaxies at high redshift that match empirical star formation laws. Yet these models  provide no mechanism to account for the  steep decrease in stellar mass of galaxies found in lower mass galaxies, even accounting for the fact that lower mass galaxies have higher gas contents \citep[e.g.][]{peeples11}.  

A contrasting model invoking energy ``feedback" from supernovae to regulate star formation has successfully simulated dwarf and Milky Way mass galaxies that share many properties of observed galaxies \citep{governato10,piontek11,guedes11}. Despite these successes, the galaxies simulated so far with the supernova energy feedback models form too many stars relative to their total mass \citep{sawala11,piontek11,avilareese11}, particularly in low mass simulations.  Additionally, the simulations fail to match the specific star formation rates of observed galaxies \citep{colin10,avilareese11}. 

In this paper we take the latter approach of directly injecting SNe energy. A clue to improving on such models came from 
measuring the velocity dispersion of the gas of our simulated dwarf 
galaxies, which was found to be significantly less turbulent than 
observed dwarf galaxies \citep{pilkington11}, suggesting that 
``feedback" energy has been under-estimated. Recent evidence suggests that radiative feedback from massive stars before they explode as supernovae can have significant effects in regulating star formation and adding energy to gas that surrounds newly forming clusters of stars \citep{nath09,murray11,hopkins11,hopkins12}.  Young, massive stars exert forces from their UV photons, stellar winds, and warm gas pressure from photo-ionized regions. In massive galaxies, momentum driven winds due to radiation pressure may be the primary driver of outflows \citep{murray05,hopkins12}. However, in galaxies of mass relevant to our study, the effect of the radiation feedback from  massive stars  is primarily  to (i) prevent collapse of gas to small, dense regions,  dissociating GMCs and  regulating star formation, and (ii)  stir up gas in star forming regions, punching holes that allow SNe remnants to expand and drive outflows \citep{hopkins12}, and (iii) provide pressure support in the disk. 

Our model does not resolve this local radiation feedback from star forming regions nor individual supernovae,  but it does resolve the shells and bubbles that are created by overlapping supernovae. Without the ability to resolve the details, we  employ a  relatively crude thermal implementation of radiation feedback from massive stars, with the  aim being to mimic their most important effects on scales that we resolve, i.e. to regulate star formation, and enhance inhomogeneity in the ISM, creating turbulence that better matches observations, and  to allow the expansion of  the SNe driven super-bubbles which drive outflows. 

Using this model, we simulate four galaxies with stellar masses ranging from 2$\times10^8$ to 1.4$\times10^{10}$M$_\odot$. Crucially, the simulated galaxies (i) were run at the same resolution with identical input physics,  and (ii) vary in their mass assembly history.  Comparing the two most massive  galaxies, SG3 has most of its merger activity at early times, prior to $z=2$, whilst in SG4 significant merging activity occurs after $z=1$.  To illustrate the effect of feedback, we include results of SG1  using less feedback (SG1LF). It forms a galaxy with too many stars and is too compact. To study the way our model varies with resolution, we re-simulated SG3 at 8 times lower resolution (SG1LR), run with the same feedback prescriptions, as well as a more massive galaxy SG5LR which extends the mass range of our sample out to 3$\times10^{10}$M$_\odot$. These lower resolution simulated galaxies also  match the scaling relations. 

Section~\ref{sec:code} describes the code and the initial conditions. Section~\ref{sec:props} presents the basic properties of the simulations including star formation histories, rotation curves, radial light and dark matter density profiles. In Section~\ref{sec:scale} we show that the galaxies all fit on the observed relations between  rotation velocity, size, luminosity, star formation rate, stellar mass, virial mass, HI mass, total baryon mass  and metallicity.  Section~\ref{sec:outflows} shows that the outflows, which are central to our models, produce metal-enriched hot halos that reproduce the observed OVI column densities in the circum-galactic-medium \citep[CGM,][]{prochaska11,tumlinson11}. In Section~\ref{sec:summary} we summarise our model for disc galaxy formation.


\renewcommand{\thefootnote}{\alph{footnote}}
\begin{table*}
\label{tab:data}
\begin{minipage}{180mm}
\begin{center}
\caption{Simulation data}
\begin{tabular}{lllllllllll}
\hline
Name & MUGS  &gas part. &IMF &$c_\star$ &M$_{halo}$ & M$_\star$  & $M_R$ & $\mu_0^*$ & S$^*$ & $n$$^{**}$ \\
         &label &mass [M$_\odot$]  &   &   & [M$_\odot$]& [M$_\odot$] & & &  &  \\
\hline
SG1             & g5664    &     2.5$\times$10$^4$  & Chabrier              &0.17 & $6.5$$\times$10$^{10}$ & $2.3$$\times$10$^{8}$ & -17.0 & 21.2 & 1.2 & 0.82\\
SG2              & g1536    &     2.5$\times$10$^4$  & Chabrier              & 0.17&$8.3$$\times$10$^{10}$ &  $4.5$$\times$10$^{8}$ & -17.5 & 21.4 & 1.4 & 0.77\\
SG3              & g15784  &     2.5$\times$10$^4$   & Chabrier              &0.17& $1.8$$\times$10$^{11}$ &  $4.2$$\times$10$^{9}$ & -20.0 & 20.0 & 2.0 & 1.17 \\
SG4               & g157807 &    2.5$\times$10$^4$ & Chabrier               &0.17&$3.2$$\times$10$^{11}$ & $1.4$$\times$10$^{10}$ & -21.2 & 19.0 & 2.5 & 1.23\\
SG1LF       & g5664     & 2.5$\times$10$^4$     & Kroupa                 &0.17 &$7.3$$\times$10$^{10}$ & $8.7$$\times$10$^{9}$  &-19.9   & 18.9 &1.0 &1.4  \\
SG3LR      & g15784    &  2.0$\times$10$^5$    & Chabrier            &0.33   &$1.8$$\times$10$^{11}$ & $4.6$$\times$10$^{9}$ & -20.1  &   19.9       &   2.1  &1.1\\
SG5LR       & g1536      &   2.0$\times$10$^5$   & Chabrier            &0.33  &$7.6$$\times$10$^{11}$ & $3.0$$\times$10$^{10}$ & -21.7  &   20.5     &     3.6    &1.3\\
\hline
\end{tabular}\\
{ $^*$central surface brightness ($\mu_o$), scale-lengths (size, S) and sersic indices ($n$) from fits in the I band\\
$^{**}$SG1-4 and SGLR are single sersic fits, SG1LF and SG5LR are 2 component sersic bulge$+$exponential disk fits} 
\end{center}
\end{minipage}
\end{table*}

\section{The Simulations}\label{sec:code}
\subsection{The simulation code: GASOLINE }

We have used {\tt GASOLINE} \citep{wadsley04}, a fully parallel, N-Body+smoothed particle hydrodynamics (SPH)
code,  to compute the evolution of the collisionless and
dissipative components in the simulations. The essential features of the code are outlined here.


Fluid elements representing gas are integrated using Smooth Particle Hydrodynamics \citep[SPH,][]{gingold77,monaghan92}. {\tt GASOLINE} is fully Lagrangian, spatially and temporally adaptive, and efficient
for large $N$. Dissipation in shocks is modelled using the
quadratic term of the standard Monaghan artificial viscosity \citep{monaghan92}.
The Balsara correction term is used to reduce unwanted shear viscosity.
{\tt GASOLINE} employs cooling due to H, He, and a variety of metal lines \citep{shen10}. The metal cooling grid is constructed using CLOUDY (version 07.02 \citealt{ferland98}), assuming ionisation equilibrium. A uniform ultraviolet ionising background \citep{haardtmadau96} is used in order to calculate the metal cooling rates self-consistently. Two mechanisms are used to prevent gas from collapsing to higher densities than SPH can physically resolve: (i) to ensure that gas resolves the Jeans mass and does not artificially fragment, pressure is added to the gas in high density star forming regions \citep{robertson08}, (ii) a maximum density limit is imposed by setting a minimum SPH smoothing length of 0.25 times that of the gravitational softening length of {$\epsilon$} = 155pc. Each simulation has between  5-15 million particles within the virial radius at z = 0, with mean stellar particle mass of $\sim$4800\,M$_\odot$ ($\sim$38000\,M$_\odot$ in the low resolution runs). 

\subsection{The Initial Conditions}

The simulations described here are cosmological zoom simulations derived from the McMaster Unbiased Galaxy Simulations (MUGS)\citep{stinson10}.  Here, those initial conditions are scaled down, so that rather than residing in a 68 Mpc cube, it is inside a cube with 34 Mpc sides. This resizing allows us to compare galaxies with exactly the same merger histories at a variety of masses. Differences in the underlying power spectrum that result from this rescaling are minor \citep{springel08,maccio08,kannan12},  and do not effect our results. SG5LR is run with the original 68 Mpc cube, having the same initial conditions as SG3.

Table 1 shows the properties of the galaxies that were selected and rescaled from the MUGS sample.  Note that SG3 uses the same initial conditions used in \cite{brook12}, which showed the impact of galactic fountains on angular momentum distribution of galaxies.  However, SG3 uses a different initial mass function (IMF) from \citep{brook12}, replacing \cite{kroupa93} with the more commonly used  \cite{chabrier03}, which is more top heavy. The conclusions of \cite{brook12} remain valid in the new simulations.

\subsection{Star Formation and Feedback}
When gas reaches cool  temperatures ($T < 10,000$ K) in a dense  environment ($n_{th} > 9.3$ cm$^{-3}$), it becomes eligible to form stars.  This value for $n_{th}$ is the maximum density gas can reach using gravity, 32 m$_{gas}/\epsilon{^3}$.  Such gas is converted to stars according to the Schmidt Law:
\begin{equation}
\frac{\Delta \rm{M}_\star}{\Delta t} = c_\star \frac{\rm{m}_{gas}}{t_{dyn}}
\end{equation}
where $\Delta$M$_\star$ is the mass of the stars formed in $\Delta t$, the time between star formation events (0.8 Myrs in these simulations), m$_{gas}$ is the mass of the gas particle, $t_{dyn}$ is the gas particle's dynamical time and $c_\star$ is the efficiency of star formation, in other words, the fraction of gas that will be converted into stars during $t_{dyn}$. Effective star formation rates are determined by the combination and interplay of  $c_\star$ and feedback, and so degeneracies do exist between feedback energy and the value of $c_\star$. In this study,  $c_\star$ is ultimately the free parameter that sets the balance of the baryon cycle off cooling gas, star formation, and gas heating. In our fiducial runs,  $c_\star$=1.67\%. 

\begin{figure*}
\includegraphics{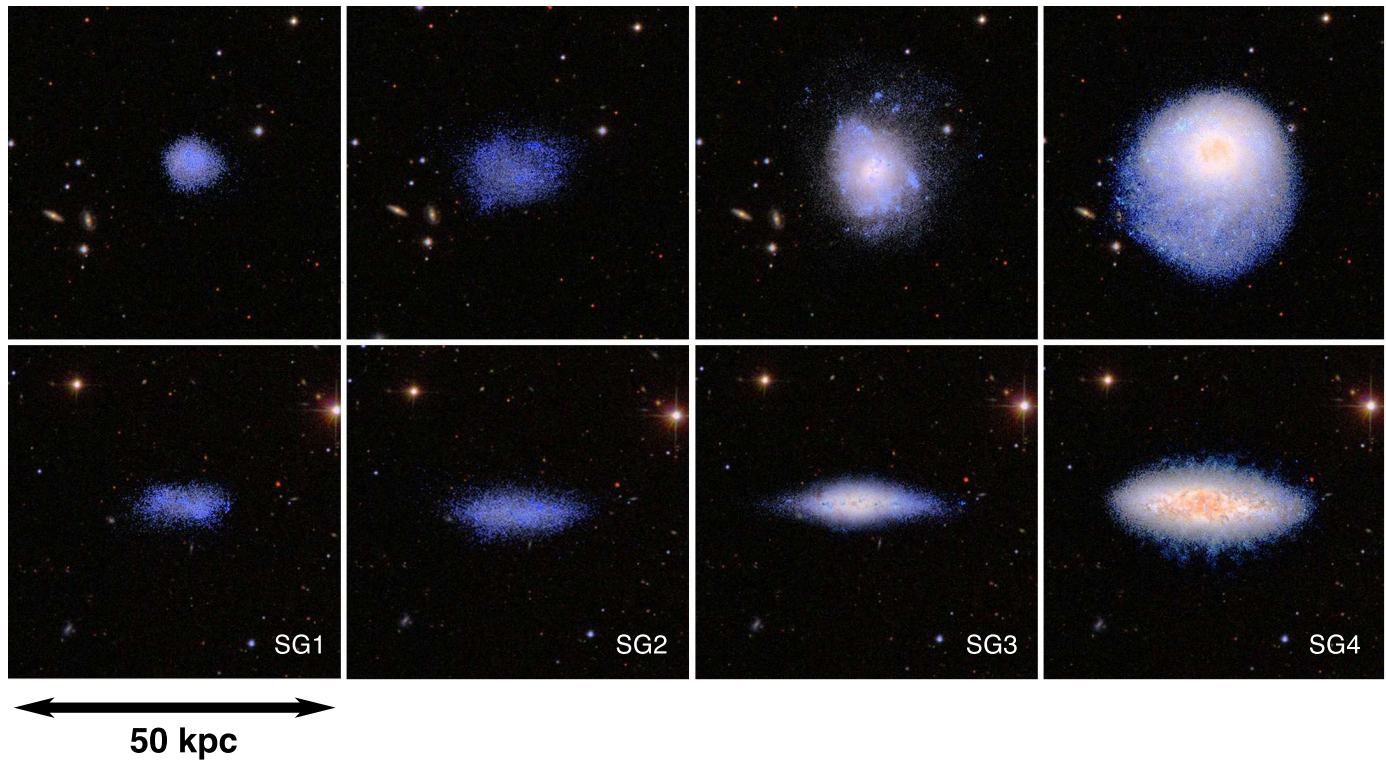}
\caption[]{\small  Mock images of the four simulated galaxies, made with {\tt SUNRISE} \citep{jonsson06} that uses the ages and metallicities of simulated star particles, and the effects of dust reprocessing, to create stellar energy distributions. Filters which mimic Sloan Digital Sky Survey (SDSS) bands i, r, and g, are assigned to colours R, G, and B respectively to produce these images. The galaxies, from left to right, SG1, SG2, SG3 and SG4,  are shown face-on (top panels)  and edge-on (bottom panels). The significant asymmetry in SG4 is caused by the late accretion of a low mass satellite galaxy.  
}
\label{fig:sdss}
\end{figure*}

Stars feed energy back into surrounding gas.  Two types of energetic feedback are considered in these simulations, supernovae and early stellar radiation  feedback from massive stars.  Supernova feedback is implemented using the blastwave formalism \citep{stinson06} and deposits $10^{51}$ erg of energy into the surrounding medium at the end of the stellar lifetime of every star more massive than 8 M$_\odot$.  Since stars form from dense gas, the energy would be quickly radiated away due to the efficient cooling.  For this reason, cooling is disabled for particles inside a blast region of size $R = 10^{1.74}E_{\rm 51}^{0.32}n_0^{-0.16}\tilde{P}_{\rm 04}^{-0.20} {\rm pc}$ and for the length of time $t = 10^{6.85}E_{\rm 51}^{0.32}n_0^{0.34}\tilde{P}_{\rm 04}^{-0.70} {\rm yr}$. Here, $E_{\rm 51}=10^{51}$ erg, $n_0$ is the ambient hydrogen density, and
${P}_{\rm 04} = 10^{-4}P_0k^{-1}$, where $P_0$ is the ambient pressure and k is the
Boltzmann constant. Both $n_0$ and $P_0$ are calculated using the SPH
kernel for the gas particles surrounding the star.

Metals are ejected from type II supernovae (SNII), type Ia supernovae
(SNIa), and the stellar winds driven from asymptotic giant branch
(AGB) stars. ÊEjected mass and  metals are distributed to the nearest neighbour gas particles using the smoothing kernel \citep{stinson06}, using literature yields for  SNII \citep{woosley95} and SNIa \citep{nomoto97}. Metal diffusion is also included, such that unresolved turbulent mixing is treated as a shear-dependent diffusion term \citep{shen10}. This allows proximate gas particles to mix their metals. Metal cooling is calculated based on the diffused metals.

Radiation energy feedback  from massive stars has been included in our model. To model the luminosity of stars, a simple fit of the mass-luminosity relationship observed in binary star systems by \citet{torres10} is used:
\begin{equation}
\frac{L}{L_\odot} = 
\begin{cases}
(\frac{M}{M_\odot})^{4},  & M < 10 M_\odot\, \\
100(\frac{M}{M_\odot})^{2},   & M > 10 M_\odot\, \\
\end{cases}
\end{equation}
Typically, this relationship leads to $2\times10^{50}$ ergs of energy being released from the high mass stars per M$_\odot$ of the entire stellar population over the 4.5 Myr between the star's formation and the commencing of SNII. These photons do not couple efficiently with the surrounding ISM \citep{freyer06}.  We thus do not want to couple all of this energy to the surrounding gas in the simulation. To mimic this highly inefficient energy coupling, we inject 10\% of the energy as thermal energy in the surrounding gas, and cooling is \emph{not} turned off for this form of energy input. It is well established that such thermal energy injection is highly inefficient at the spatial and temporal resolution of  the type of cosmological simulations used here \citep{katz92,kay02}. This is primarily due to the characteristic cooling timescales in the star forming regions being lower than the dynamical time. 

In star forming regions where gas has density of $\sim 10$ cm$^{-3}$, early stellar feedback typically heats the surrounding gas to  between $\sim 1$ to a few $\times 10^6$K.  At the density where stars form in our simulations,  i.e. n$\sim 10$ cm$^{-3}$, cooling times are significantly shorter than dynamical times, even for temperatures of order $10^6$K  (see e.g. Figure\,1 in \citealt{dalla12}), typical of those which to our gas is heated by our radiation energy feedback, meaning that it is only gas particles which escape to less dense regions that  have any affect from this feedback. We found that only $\sim$10\% of heated particles had dynamical times longer than cooling times at high redshift (3$>$z$>$1) when the ISM is most turbulent and the metallicity is low, reducing to only $\sim$3\% by $z=0$. Thus, between $90-97$\% is radiated away within a single dynamical time, meaning that our effective efficiency of coupling radiation energy feedback to the ISM is between $0.3-$1\%. It also means that our implementation does not evenly heat gas, i.e. we do not effectively couple the $\lesssim$1\% of energy to all gas particles affected by radiation energy. Rather, our scheme is essentially stochastic, with a small number of gas particles effected. This scheme is thus somewhat similar to the SNe feedback implementation of \cite{dalla12}.   This radiation energy feedback does not directly drive outflows:  implemented in the absence of SNe feedback results in no outflows.  Rather, it helps to  regulate star formation, and enhances inhomogeneity in the ISM, creating increased turbulence,  and allowing the expansion of the SNe driven super-bubbles which drive outflows.

To understand the effect of the strengthened feedback, a  version of SG1 was simulated using the stellar feedback as implemented in \citet{stinson10} so that stars form when gas reached 1 cm$^{-3}$, with a \citet{kroupa93} IMF, with 0.4$\times 10^{51}$ ergs deposited per supernova explosion. We refer to this as the Òlower feedback modelÓ (LF) in this study, but note that this feedback strength is comparable to  most implementations that are currently run in the literature \citep{scannapieco12}.

\subsection{Resolution Dependence}
To study the resolution dependence of our model, we also simulated two galaxies at 8$\times$ lower resolution: SG3LR and SG5LR. Clearly, SG3LR is the low resolution version of SG3, while SG5LR is a galaxy with stellar (total) mass of 3.0$\times10^{10}$\,(7.6$\times 10^{11}$)\,M$_\odot$. As with other galaxy formation simulations in the literature, galaxy properties are not precisely the same at different resolutions when the same parameters are used \cite[e.g.][]{scannapieco12}. In this study we aim to retain the same baryon cycle at the two different resolutions as this drives the simulated galaxy properties. Hence we want our star formation rates to match at the two resolutions. 
 To achieve this  we adjust our free parameter c$_\star$, the input star formation efficiency, to ensure that the star formation rate remains the same at the two resolutions.   With c$_\star$=0.033, SG3LR has a star formation rate that closely matches our fiducial high resolution run, SG3. 

As the feedback implementation is the same between the two resolutions, the resultant galaxies have the same balance of star formation and feedback, and hence very similar baryon cycles, meaning that same key processes occur on the scales that we resolve i.e. the same  the baryon cycle between star formation, hot and cold gas, outflows and the density of gas in the star forming regions. At z=0,  total baryon mass within the virial radius,  the mass of stars, mass of  hot gas ($>$4$\times$10$^4$K), and cold gas ($<$4$\times$10$^4$K) in SG3LR are all within 5-10\% of the values for SG3. We will see that the resultant galaxies have almost identical properties with respect to scaling relations. We subsequently use the same calibration and input parameters for SG5LR, a more massive galaxy. This shows that we can form reasonable galaxies even at relatively low resolution, and it allows us to increase the range of masses over which our simulated galaxies match the scaling relations.

\section{Simulated Galaxy Characteristics}\label{sec:props}
Mock observations of the four main simulated galaxies are shown in Figure~\ref{fig:sdss}.  These images were created using the Monte Carlo radiative transfer code {\tt SUNRISE} \citep{jonsson06}, which simulates the effect of dust absorption.  The morphologies of the galaxies range from  dwarf irregular (SG1, SG2) to more structured disc galaxies (SG3, SG4). The significant asymmetry in SG4 is caused by the late 
accretion of a low mass satellite galaxy.
 \subsection{Star Formation and Mass Accretion}
\begin{figure}
\hspace{-0.5cm}\includegraphics[height=0.25\textheight]{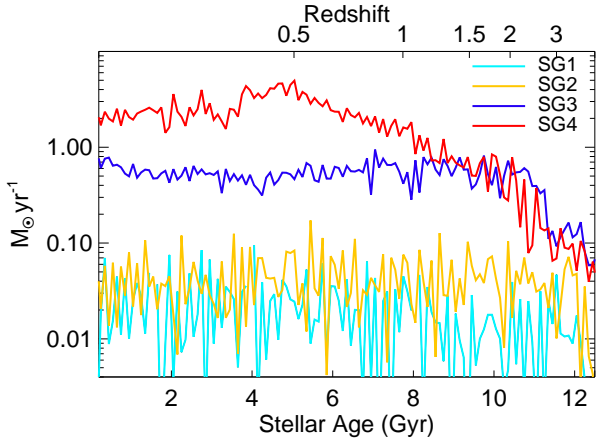}
\caption{The star formation histories of the four simulated galaxies, SG1 (cyan line), SG2 (yellow), SG3 (blue), and SG4 (red).  SG1 and SG2  have a bursty, yet reasonably consistent star formation histories. SG3 has a peak at high redshift, when it has its  merging epoch, while SG4 has a much later merging epoch, resulting in a star formation rate that peaks at a later time.}
\label{fig:sfr}
\end{figure}

\begin{figure}
\hspace{-0.5cm}\includegraphics[height=0.25\textheight]{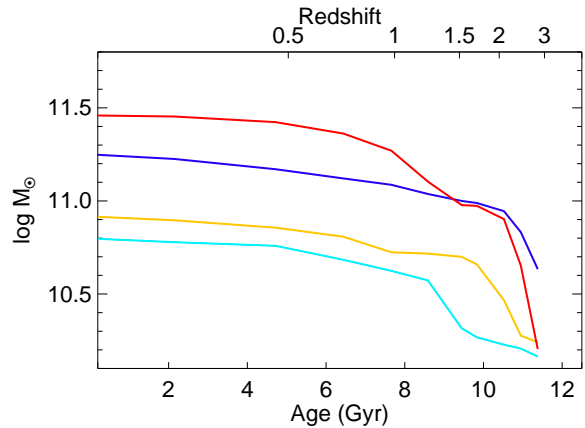}
\caption{The growth of the virial mass of the halos of the simulated galaxies.  The later build up of mass in SG4, which is still undergoing significant merging activity well after z=1,  relates directly to its later peak in star formation. }
\label{fig:massgrowth}
\end{figure}

Figure~\ref{fig:sfr} shows the star formation histories of the simulated galaxies, SG1 (cyan line), SG2 (yellow), SG3 (blue) and SG4 (red). SG1 and SG2 both have  bursty, yet reasonably consistent star formation histories. SG3 has a peak at high redshift, when it has its  merging epoch, while SG4 has a much later merging epoch, resulting  a star formation rate that peaks at a later time.  The growth of the virial mass of the halos of the simulated galaxies is shown in  Figure~\ref{fig:massgrowth}.   Merger events appear as significant increases in the total mass. The later build-up of mass in SG4 relates directly to its later peak in star formation.

\subsection{Surface Brightness Profiles}
\begin{figure*} 
\hspace{-0.4cm}\includegraphics[height=.18\textheight]{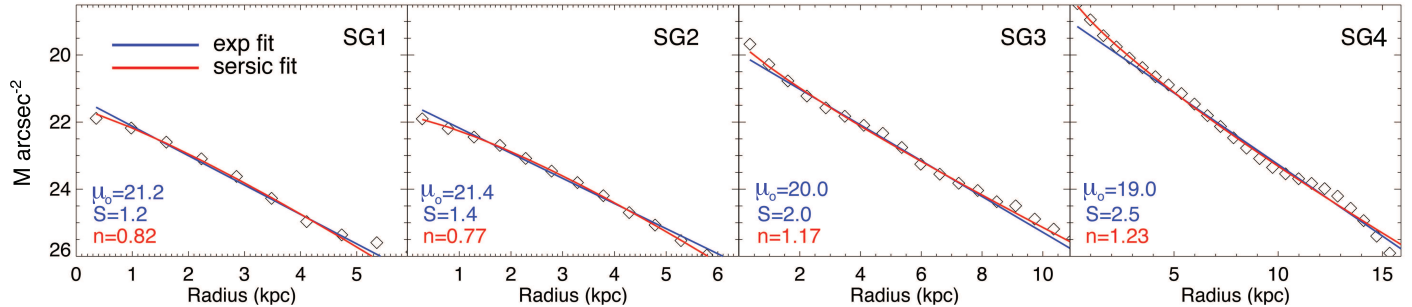}
\caption{The I band surface brightness profiles of SG1, SG2, SG3 and SG4 from left to right respectively. Fits are made using both exponential (blue lines) and sersic (red lines) profiles. Central surface brightness ($\mu_o$) and scale-lengths (S) of the exponential fits as well as  the sersic index ($n$) from the sersic fits are noted in the bottom left of each panel. In each case fits are made from radius=0 out to  I=25\,M\,arcsec$^{-2}$.}
 \label{fig:profile}
\end{figure*}

Figure~\ref{fig:profile} shows exponential (blue line) and single sersic (red line) fits to the I band profiles of the face-on surface brightness maps of each simulated galaxy, measured from face-on images and including the effects of dust reprocessing using {\tt SUNRISE}. In each case we fit from the centre out to the radius where the I band surface brightness  is 25\,M\,arcsec$^{-2}$. The central surface brightness ($\mu_o$), exponential scale-length (S),  and sersic index ($n$) are shown. The values of scale-length (S) are used for the ``size" of the galaxy in this paper, to make consistent with the observational data with which comparison is made \citep{courteau07}. Each galaxy exhibits a nearly exponential surface brightness profile.  In every case, the Sersic index, $n<1.5$.

\subsection{Rotation Curves}
\begin{figure}
\hspace{-0.3cm}\includegraphics[height=.2\textheight]{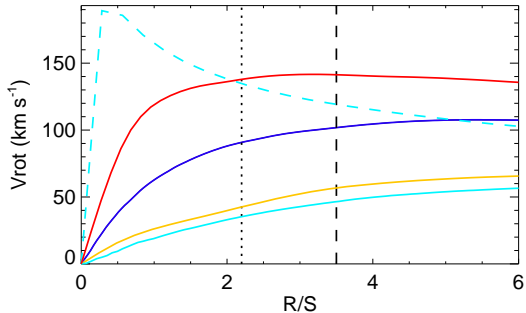}
\caption{Rotation velocity (V$_{\rm c}$) versus radius of SG1 (cyan line), SG2 (yellow), SG3 (blue) and SG4 (red).   Radial units are scaled by the disc scale-length in each case. The low feedback simulation, SG1LF (dashed cyan line) shows the extreme difference in stars formed compared to the fiducial run, particularly in the central regions, combined with adiabatic contraction rather than expansion.  The dotted and dashed vertical lines are at 2.2 and 3.5 scale-lengths. In the paper, we use V$_{\rm c}$ as measured at 3.5 scale-lengths in each case, as it  better reflects the underlying mass of the system. }
\label{fig:rotcur}
\end{figure}

Figure~\ref{fig:rotcur} shows the rotation curves of four main galaxies along with the low feedback version of SG1. The rotation  velocities (V$_{\rm c}$) are computed by calculating the mass within spherical radial shells $M(r)$, then relating via V$_c(r)=\sqrt{(GM(r)/r)}$ where G is Newton's gravitational constant.   Radial units are scaled by the disc scale-length in each case. The excessive central mass distribution that has plagued simulations of galaxy formation \citep{mayer08} is absent in each case, reflected in the lack of  central ``peak" in the rotation curves. The vertical dotted and dashed lines are at 2.2 and 3.5 scale-lengths, where observational determinations of rotational velocity are often made \citep{courteau97,giovanelli97a}.  We can see that in SG1, the rotation curve is still rising quite rapidly at 2.2 scale-lengths. This  issue of rising rotation curves is also common in observations of low mass galaxies \citep{epinat08}. In this paper, we report the value of V$_{\rm c}$ at 3.5 scale-lengths in each case, as it  best reflects the underlying mass of the system.  Also shown is SG1LF, as a dashed cyan  line. The contrast between the two SG1 galaxies is extreme, with much of the available baryons forming stars in the central region of the low feedback run, making the central region very dense and the rotation velocity very high in this region.

\subsection{Dark Matter Profiles: Cores not Cusps}
\begin{figure}
\hspace{-0.3cm}\includegraphics[height=.2\textheight]{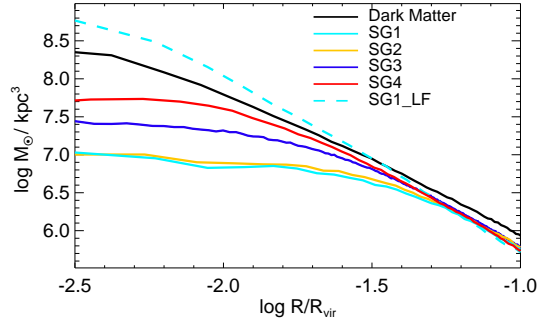}
\caption{The dark matter density profiles, on  a log-log scale, where radial units are  scaled by the virial radius. Also plotted is a dark matter only simulation of SG3, which has the  steep inner profile that is characteristic of dark matter halos. The low feedback simulation SG1LF (dashed cyan line)  shows significant adiabatic contraction. 
}
\label{fig:core}
\end{figure}

The dark matter density profile, on  a log-log scale, is plotted for SG1 (cyan line), SG2 (yellow), SG3 (blue), and SG4 (red) in  Figure~\ref{fig:core}. Also plotted is a dark matter only simulation of SG3, which has the  steep inner profile that is characteristic of dark matter halos \citep{navarro95}. Radial units are scaled by the virial radius in each case. In all the simulated galaxies, the dark matter profile is less steep than the profiles in pure dark matter simulations. This flattening of the central dark matter profile  when baryons are included has also been found in cosmological simulations of dwarf galaxies \citep{governato10} and massive disc galaxies \citep{maccio12}. This contrasts with simulations that do not include strong feedback, which invariably have dark matter profiles that are even steeper than in pure dark matter simulations \citep{gnedin11} because the dark matter adiabatically contracts as gas cools to the center.  SG1LF (dashed cyan line) shows exactly this behaviour as its dark matter density profile is significantly steeper than the simulation that only includes dark matter.   The fact that the amount of the outflows from the central region in the present study results in matching the scaling relations,  favours the findings of simulations where ``expansion" rather than contraction occurs over the masses of these simulations, as semi-analytic models have predicted \citep{dutton09}.

\section{Scaling relations}\label{sec:scale}
The crucial test of galaxy formation models does not come from matching a single relation, as that can often be achieved through a parameter search of input physics. Rather, models are judged by their ability to simultaneously match a range of observational constraints.  

\subsection{Definitions}
\emph{ Halo Mass}:  To make a direct comparison to the abundance matching results, we use $M_{200}$ from our dark matter only runs.  As described in \citet{stinson10}, one step in creating the initial conditions for the MUGS galaxies was running a zoom dark matter only simulation at the same resolution as the simulation that includes baryons.  $M_{200}$ is the mass enclosed inside $r_{200}$, the radius at which the density reaches  200$\rho_c$, where $\rho_c$ is the critical density.\\
\emph{Stellar Mass:}  The sum of the mass of all stars within the radius where the I band surface brightness  is 25 M\,arcsec$^{-2}$.\\
\emph{HI mass:}  The Saha equation is solved to determine an ionization equilibrium. This remains an approximation since an accurate model of HI mass would require full radiative transfer included in the code and is beyond the scope of this paper. In particular self shielding from the UV background is not included in our model and may result in our derived HI masses being under estimates, while  photo-ionization of HI from the galaxy itself it also excluded. \\
\emph{Cold gas mass:}  4/3 $\times$$M_{HI}(5 h_r)$.  That is 4/3 the HI mass contained within five scale-lengths of the galaxy.  A factor between 1.2 and 1.4 is commonly used in observations to account for heavy elements.  We use 4/3 as it is used in the relevant observational studies to which we compare \citep{mcgaugh05}.\\
\emph{Magnitudes, luminosities and colours}: come directly from the {\tt SUNRISE }  outputs, and include the affects of dust reprocessing on the spectral energy distribution.

\subsection{Relationships}
\begin{figure}
\hspace{-0.35cm}\includegraphics[height=.375\textheight]{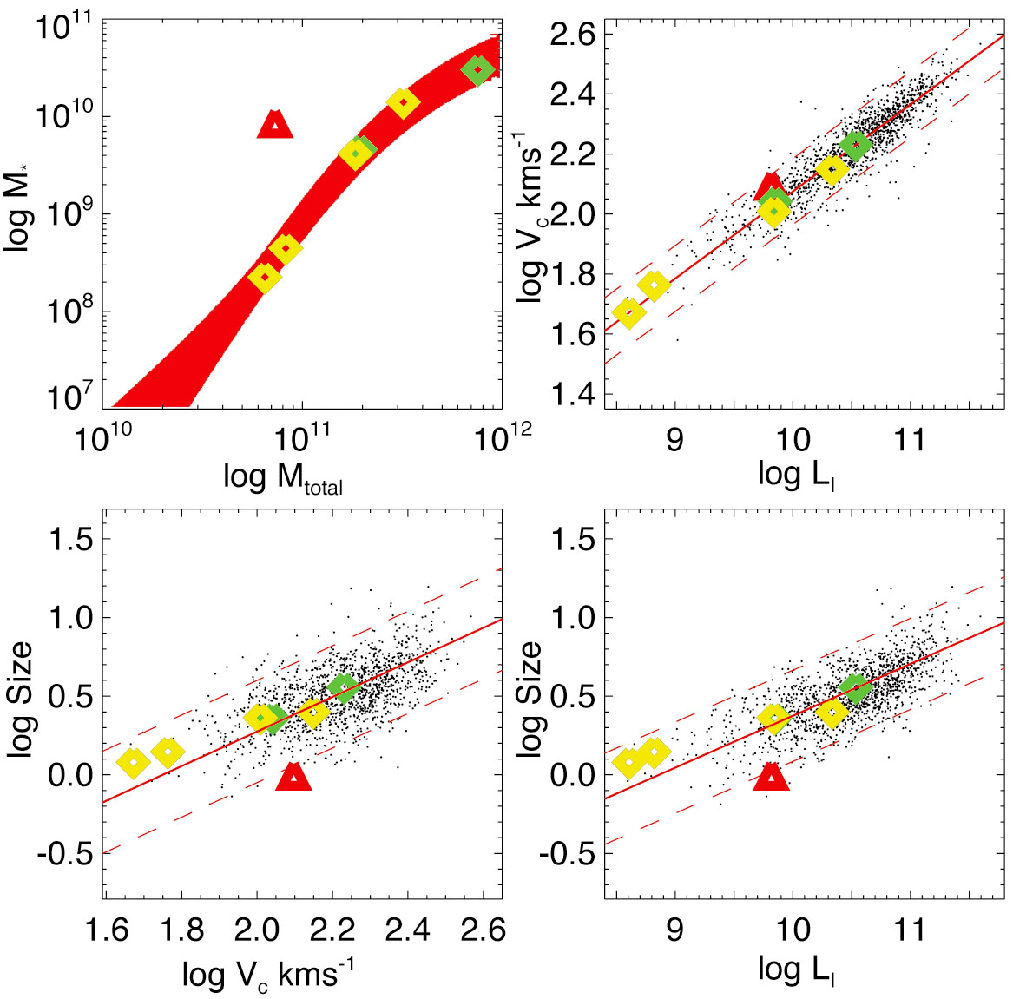}
\caption{Scaling relations of the simulated galaxies  overplotted on observational data. All plots are shown on a log scale. Panel (a) Stellar mass (M$_*$) against total mass (M$_{\rm vir}$). The thick red line shows the (semi) empirical relation \citep{sawala11}. All other plots use a single  observational data set \citep{courteau07}. Red lines are fits to the observational data while dashed red lines include 97\% of data.   Panel (b) Rotational velocity (V$_{\rm c}$) against  Luminosity (L$_{\rm I}$) in the I-band (the Tully-Fisher relation). Panel (c) Size (S), or disc scale-length, against L$_{\rm I}$. Panel (d) S against V$_{\rm c}$.   The four fiducial runs are shown as yellow diamonds, the low resolution runs as green diamonds and the low feedback run as a red triangle.}
\label{fig:scaling1}
\end{figure}

Panel\,(a) of Figure~\ref{fig:scaling1} shows the relationship between the stellar mass and host halo mass of the four fiducial simulated galaxies as yellow diamonds, over-plotted on the published \citet{guo10} relationship.  Our free parameter, $c_\star$, was tuned to place SG3 on this relationship.  There was no guarantee that the other simulated galaxies should fall on this  relationship.  The red triangle represents the low feedback case, SG1LF, which demonstrates that low feedback results in  more than an order of magnitude too many stars. 

The green diamonds represents the low resolution runs. As stated, we adjusted   $c_\star$ such that the star formation rate in SG3LR matched as closely as possible to that of SG3.   As we have not changed our feedback recipe, the balance between star formation and energy feedback remains the same at the two resolutions. What is interesting is that using these parameters for a more massive galaxy,   SG5LR, the simulation again fits on the relation, extending the mass range over which our balance between star formation and feedback results in the correct scaling of stellar mass with halo mass.  A more thorough analysis of SG5LR will be made in a forthcoming study.

\begin{figure}
\hspace{-0.35cm}\includegraphics[height=.375\textheight]{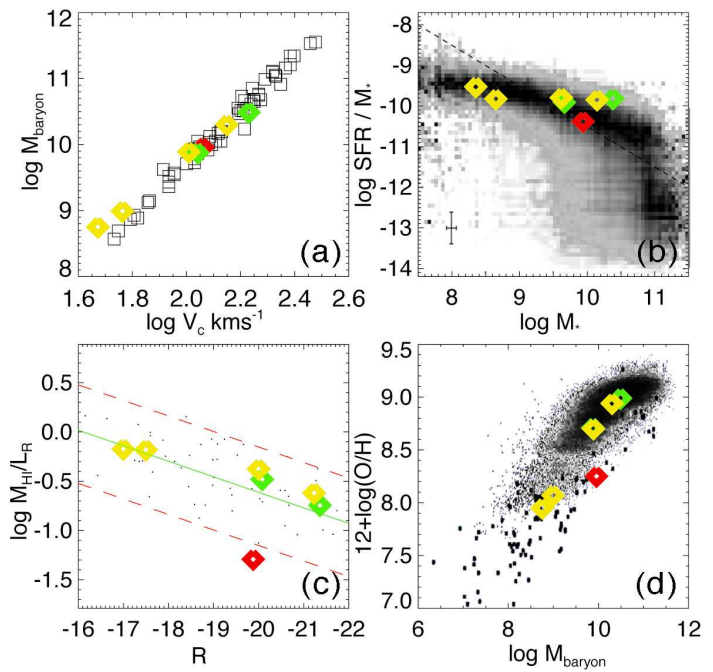}
\caption{Simulated galaxies overplotted on observational data.  Panel (a)  Total baryonic mass (M$_{\rm baryon}$), consisting of all stars and cold gas, plotted against  V$_{\rm c}$ \citep[observed data from][]{mcgaugh05}.    Panel (b) Specific star formation rate against  stellar mass  \citep[observed data from][]{salim07}.  Panel (c) The ratio of the mass of neutral hydrogen gas (H\,I) to luminosity in the R-band (L$_{\rm R}$) plotted against the R-band magnitude \citep[observed data from][]{verheijen01}. Red lines are fits to the observational data while dashed red lines include 97\% of data.   Panel (d) The ratio of Oxygen to Hydrogen, 12+log(O/H), against total baryonic mass  \citep[observed data from][]{tremonti04}. The four fiducial runs are shown as yellow diamonds, the low resolution runs as green diamonds and the low feedback run as a red triangle.}
\label{fig:scaling2}
\end{figure}

Figures\,\ref{fig:scaling1}\,\&\,\ref{fig:scaling2} show the relationships between the rotation velocity, size, luminosity, star formation rate stellar mass, total  mass, HI mass, total baryon mass and metallicity of the simulated disc galaxies. The fiducial simulations match the relations remarkably, and their properties correctly scale over the mass range of the simulations. Again, we emphasise that none of the parameters were tuned in order to match any of these relations. As stated, the only tuning was to adjust $c_\star$ such that SG3 (and SG3LR) fits on the stellar mass-halo mass relation.  The low resolution cases (green diamonds) also match the relations, with SG3LR matching well its high resolution counterpart. 

It is very interesting that the thermal feedback implementation results in stellar mass-halo mass relation that has such steep dependence on mass (panel 7a). Our feedback has no mass dependence nor mass loading included in its implementation. It simply inputs a given amount of energy per unit of star formation, resulting in  pressure from thermal energy that drives outflows, controlling the supply of cold gas available for star formation. Feedback processes then further regulate the star formation rates of this cold gas. 

The other panels in Figure~7 (b-d) relate to the angular momentum of the galaxies at given luminosities.  Matching the Tully-Fisher relation over a wide range in luminosities, combined with making disk galaxies of the correct size (without large stellar bulges), indicates that we have attained the correct amount  and distribution of angular momentum. The angular momentum of the baryons in these simulations do not simply follow the dark matter, as analytic models often assume. Rather, the distributions arise from a cycle of	 cooling to the star forming regions,  outflows, and galactic fountains \cite[see][]{brook11,brook12}.

The relations in Figure~8 put  further constraints on the baryon cycle of galaxies of varying mass.  Even though our feedback drives outflows, it does not simply eject all the cold gas from the disk. This is the result of the inhomogeneous interstellar medium, with pockets of hot gas building pressure and driving outflows. This is key in allowing the high HI fractions  in the  low mass galaxies (Panel 8c), even though they  lose a high fraction of their baryons  in the outflows. These processes also relate directly to the success of the low mass simulations in matching the specific star formation rates.  Simulations have largely had trouble  matching the specific star formation rate for low mass galaxies \citep{colin10,avilareese11}. With low feedback, the star formation rates follow the gas accretion rates, which exponentially decline after $z\sim 2$, meaning that the stars form too early. Attempts to form significantly fewer stars in dwarf galaxies by increasing feedback had the trouble of ejecting their gas in high   redshift star bursts, resulting in isolated dwarf spheroidal galaxies \citep{sawala10}, rather than dwarfs with prolonged star formation histories and significant amounts of HI gas at $z=0$, as observed in the field \cite[e.g.][]{geha06}. Our feedback implementation is able to maintain significant amounts of cold gas in the star forming region, to regulate star formation whilst driving outflows and gas recycling via galactic fountains. 

The match to the baryonic mass-metallicity relation  (panel 8d)  provides  added confidence in the baryon cycle that our feedback creates. The balance between fresh gas accretion, metal injection from star forming regions, metal enriched outflows and recycling conspires to result in the correct gas metallicities at each mass range. We show in Section~5 that this  cycle of metals also results in OVI abundances in the circum-galactic medium (CGM) that match observations. Further detailed studies of  the metallicities and their distributions   will be made in forthcoming studies, placing greater constraints on our baryon cycle. 

The low feedback case (SG1LF) is also particularly interesting. The huge increase in number of stars formed means much more mass in the inner region, so the rotation velocity (measured at 3.5 scale-lengths) is much higher than in the high feedback case. Yet the increased stellar mass also results in higher luminosity:  the values conspire so that it remains on the Tully-Fisher relation. Of course, the peaked rotation curve means a there is dependence on the radius at which rotation is measured. Beyond the mass-metallicity relation, it requires the size-metallicity relation to break the degeneracy of the high and low feedback cases both fitting on the Tully-Fisher relation.  The low feedback run is too concentrated.  Thus the shape of the rotation curve and the three parameters size-luminiosity-rotation are required to assess simulations. This was pointed out in \cite{dutton08}. The low feedback run is also  deficient in HI relative to its R-band luminosity (panel 8c), due to turning too much of its cold gas into stars. 

\section{Metal Enriched Gaseous Corona}
\label{sec:outflows}

\begin{figure}
\hspace{-0.4cm}\includegraphics[height=.38\textheight]{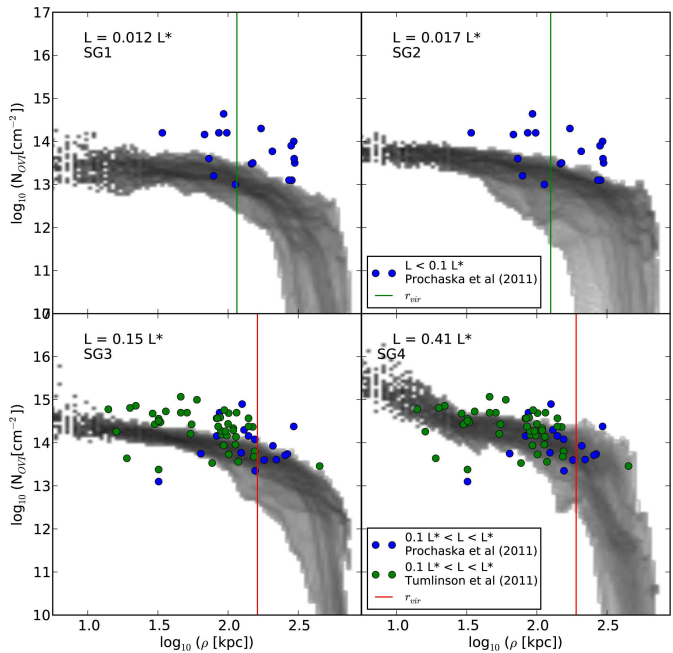}
\caption{Radial profiles of the column density maps of OVI for the 4 simulated galaxies. The large dots are observations of L $<$ 0.1 L$_*$ galaxies from  Prochaska et al. (2011) (blue, top panels)  and in the lower panels the 0.1 L$_*<$L$<$L galaxies from Prochaska et al. (2011, blue) and Tumlinson et al. (2011, green) galaxy samples. The solid vertical green (red) line represents the virial radius in the upper (lower) panels  for each of the four halos.}
\label{fig:ovi}
\end{figure}

One way to test the extent of  outflows, and hence the feedback in our model,  is  to compare the metal enriched gas surrounding the simulated galaxies with observations. \cite{prochaska11} and \cite{tumlinson11}  find that column densities of OVI extend out to 300 kpc  from star forming galaxies.  Their observations extend to galaxies as faint as 0.01 L$^\star$, making them especially useful to compare with our models.
Figure~\ref{fig:ovi} shows the surface density profiles of OVI gas as a function of the impact parameter, $\rho$, from the centre of the simulated galaxies.  The observed OVI column densities for sub-$L^*$ galaxies
at $z\sim 0$ from \cite[][blue dots]{prochaska11} and \cite[][green dots]{tumlinson11} are shown overplotted on the model data.  Each panel is labeled with the 
V-band luminosity of the simulated galaxies and the virial radius, indicated 
by the vertical green lines. Our feedback implementation produces extended, metal enriched gaseous coronae that extend to impact parameters of $\sim$300 kpc, even around low mass  halos,  matching observed absorption line features. We refer the reader to \cite{stinson12} which  presents full details of our comparisons between simulations and observations and shows that the low feedback case is significantly deficient in  OVI within its CGM.  Note that  the affects of metal diffusion on the distribution of metals were explored in \cite{shen10} and they find that the effect is minimal these results.

\section{Conclusions}\label{sec:summary}

We use  the same physical model, including feedback implementation, at the same resolution to simulate four galaxies over a wide mass range that match a  range of galaxy properties and scaling relations.  Our model includes radiation energy from massive stars, as well as supernova energy. We adjusted the star formation efficiency parameter $c_\star$ so that a single simulated galaxy, SG3, matched the stellar mass-halo mass relation \citep{moster10,guo10}. 
 With the additional energy sources  driving outflows from star formation regions in our updated model, the complex interplay between cooling gas, star formation, radiation energy injection from massive stars,  stellar winds and supernova explosions, outflows, and the recycling of gas  (the baryon cycle)  have been balanced in our model to (i) retain  the correct amount of baryons, blowing out large amounts of gas from the central regions, particularly in low mass galaxies, 
 (ii) reproduce star formation rates and histories to attain the correct galaxy luminosities, stellar masses and colours for a given mass galaxy, (iii) ensure that low mass galaxies are gas rich compared to high mass galaxies, relative to their stellar population, even though  significantly more gas is blown out of the low mass galaxies, (iv) attain an  amount and distribution of angular momentum in the star forming gas to result in the correct morphology and disc sizes, and (v) correctly balance inflows of fresh gas with the recycling of  gas  from star forming regions, in order  to correctly match the mass-metallicity relation.   
 
 We have previously shown that the outflows in our model  remove low angular momentum gas \citep{brook11}, resolving the ``angular momentum problem" which has plagued simulations \citep[e.g.][]{navarrosteinmetz00,piontek11}.  Some of the heated gas is ejected from the galaxy entirely, particularly at early times and from the low mass galaxies, while some of the gas is blown only as far as the  dark matter halo that surrounds the galaxy, and may cool back down to the star forming region at later times and form disc stars \citep{brook12}.  In our four simulated disc galaxies, outflows from the central regions also drive an  expansion of the central dark matter profiles, resulting in   flat dark matter central density profiles which better match observed galaxies, rather than the very steep density profile that are predicted from  dark matter simulations even in our most massive disc galaxy simulation. 
 
 By re-calibrating the input star formation efficiency in order to attain the same star formation rate in a lower resolution simulation (SG3LR) as in the fiducial run (SG3) without changing the feedback implementation, we were able to attain a very similar baron cycle and hence a galaxy with very similar  properties at two resolutions. Our calibration for this lower resolution also resulted in an L$_*$ galaxy (SG5LR) that matches the scaling relations. By contrast, a low feedback run (SG1LF) was shown to form too many stars, particularly in the central regions. Our study shows that it is necessary to examine a number of relations to determine the success of simulations, and we recommend plotting at minimum  the size-luminosity-rotation velocity properties rather than simply the Tully-Fisher relation (see also \citealt{dutton08}).



Our model suggests that large scale outflows of gas from the central regions of galaxies are the primary driver of galaxy scaling relations, as well as shaping disc morphology. The model makes the assumption that radiation energy from massive stars regulates star formation and   stir up gas in star forming regions, allowing SNe remnants to expand and drive outflows.  We show that the scale of outflows invoked in our models  matches the observed absorption line features of local galaxies \citep{prochaska11,tumlinson11}. A detailed comparison between the simulations and these observations is made in \cite{stinson12}, where we show that lower feedback models,  more akin to those commonly used \citep[e.g.][]{scannapieco12}, are significantly deficient in extended OVI, particularly in low mass galaxies.    

 Only in a very narrow mass range does the gravitation of host dark matter halos  balance the processes causing these  outflows, allowing disc galaxies to form. In the lower mass halos, the significant energy feedback drives increased turbulence, leading to more irregular galaxy morphologies (SG1, SG2).

\section*{Acknowledgments}

CB and BKG acknowledge the support of the UKÕs Science \& Technology Facilities Council (STFC Grant ST/F002432/1). TQ was supported by NSF grant AST- 0908499. We acknowledge the computational support provided by the UKÕs National Cosmology Supercomputer, COSMOS.
\bibliographystyle{mn2e}
\bibliography{brook}

\end{document}